\documentclass[12 pt,twoside,aps,prd,amsmath,amssymb,
tightenlines,showpacs,showkeys,eqsecnum]{revtex4-1}

\usepackage[pagebackref]{hyperref}

\usepackage{graphicx}
\usepackage{mathptmx}
\usepackage{bm}
\newcommand {\cG}{\cal G}

\newcommand {\bp}{\bar \psi}
\newcommand {\p}{\psi}

\def\myfigure #1#2#3#4
{\begin{figure}[ht]\begin{center}
\includegraphics[width=#2 \textwidth]{#1.eps}
\parbox[t]{#4\textwidth}{\caption{#3}\label{#1}}
\end{center}\end{figure}}

\def \myfigures #1#2#3#4#5#6#7#8
{\begin{figure}[ht]
    \begin{center}
        \includegraphics[width=#2 \textwidth]{#1.eps}
        \hfill
        \includegraphics[width=#6 \textwidth]{#5.eps}
        \parbox[t]{#4\textwidth}{\caption {#3}\label{#1}}
        \hfill
        \parbox[t]{#8\textwidth}{\caption {#7}\label{#5}}
    \end{center}
\end{figure} }

\begin{document}
\baselineskip -24pt
\title{Nonlinear Spinor Fields in LRS Bianchi type-I spacetime:  Theory and observation}
\author{Bijan Saha and Victor S. Rikhvitsky}
\affiliation{Laboratory of Information Technologies\\
Joint Institute for Nuclear Research\\
141980 Dubna, Moscow region, Russia} \email{bijan@jinr.ru}
\homepage{http://bijansaha.narod.ru}

\begin{abstract}

Within the scope of a LRS Bianchi type-I cosmological model we study
the role of the nonlinear spinor field in the evolution of the
Universe. In doing so we consider a polynomial type of nonlinearity
that describes different stages of the evolution. Finally we also
use the observational data to fix the problem parameters that match
best with the real picture of the evolution. The assessment of the
age of the Universe in case of the soft beginning of expansion
(initial speed of expansion in a point of singularity is equal to
zero) the age was found 15 billion years, whereas in case of the
hard beginning (nontrivial initial speed) it was found that the
Universe  is 13.7 billion years old.

\end{abstract}



\maketitle

\bigskip

\section{Introduction}

The discovery and further confirmation of the existence of the
accelerated mode of expansion of the present day Universe
\cite{riess,perlmutter} lead cosmologists to construct new theories
able to explain this new cosmological findings. Though cosmological
constant, quintessence, Chaplygin gas etc. are the prime candidates,
thanks to a number of remarkable works
\cite{henneaux,ochs,saha1997a,saha1997b,saha2001a,saha2004a,
saha2004b,saha2006c,saha2006e,saha2007,saha2006d,greene,ribas,souza,kremer},
recently many authors considered the spinor fields as a possible
alternative to these models. And it is because of the spinor fields'
ability to simulate different type of source fields from ekpyrotic
matter to phantom matter and Chaplygin gas
\cite{krechet,saha2010a,saha2010b,saha2011,saha2012}. Moreover it
was found that a nonlinear spinor field can also (i) generate
singularity-free Universe
\cite{saha1997a,saha1997b,saha2001a,saha2004a,saha2004b}; (ii)
accelerate the isotropization process of the initially anisotropic
spacetime \cite{saha2001a,saha2004a,saha2006c} and (iii) give rise
to a late time accelerated mode of expansion \cite{ribas,
saha2006d,saha2006e,saha2007}.

Some recent studies show that the non-diagonal components of the
energy-momentum tensor of the spinor field can play significant role
on the geometry of spacetime as well as on the components of the
spinor field itself, namely on the spinor field nonlinearity
\cite{FabIJTP,ELKO,FabJMP,sahaIJTP2014,sahaAPSS2015,sahabvi0,sahabvi,sahalrsbi}.
In those papers it was shown that depending on the specificity of
metric in some cases the spinor field nonlinearity and the mass
terms vanish all together, whereas in other cases both the mass term
and nonlinear term do not only disappear but also play a crucial
role in the evolution of the Universe. In a recent paper
\cite{sahalrsbi} it was found that within the scope of a LRS Bianchi
type-I spacetime the spinor field nonlinearity depending on the sign
of self-coupling constant allows either an expansion with
acceleration or an oscillatory mode of evolution of the Universe. In
this note we develop that theoretical work further and fix the
values of the problem parameters using the recent observational
data.

\section{Basic equations}

The LRS Bianchi type-I (BI) model is the ordinary Bianchi type-I
model with two of the three metric functions being equal to each
other and can be given by
\begin{equation}
ds^2 = dt^2 - a_1^{2} \left[dx^{2}\, + \,dy^{2}\right] -
a_3^{2}\,dz^2, \label{lrsbi}
\end{equation}
with $a_1$ and $a_3$ being the functions of time only.

Keeping this in mind the symmetry between $\p$ and $\bp$ we choose
the symmetrized Lagrangian \cite{kibble} for the spinor field as
\cite{saha2001a}:
\begin{equation}
L = \frac{\imath}{2} \biggl[\bp \gamma^{\mu} \nabla_{\mu} \psi-
\nabla_{\mu} \bar \psi \gamma^{\mu} \psi \biggr] - m_{\rm sp} \bp
\psi - F, \label{lspin}
\end{equation}
where the nonlinear term $F$ describes the self-interaction of a
spinor field and can be presented as some arbitrary functions of
invariants generated from the real bilinear forms of a spinor field.
We consider $F = F(K)$, with $K = \{I,\,J,\,I+J,\,I-J\}$. It can be
shown that such a choice describes the nonlinearity in its most
general form.

Varying \eqref{lspin} with respect to $\bp (\psi)$ one finds the
spinor field equations:
\begin{subequations}
\label{speq}
\begin{eqnarray}
\imath\gamma^\mu \nabla_\mu \psi - m_{\rm sp} \psi - 2 F_K (S K_I +
 \imath P K_J \gamma^5) \psi &=&0, \label{speq1} \\
\imath \nabla_\mu \bp \gamma^\mu +  m_{\rm sp} \bp + 2 F_K \bp(S K_I
+  \imath P K_J \gamma^5) &=& 0. \label{speq2}
\end{eqnarray}
\end{subequations}
Here we denote $F_K = dF/dK$, $K_I = dK/dI$ and $K_J = dK/dJ.$

The energy-momentum tensor of the spinor field is given by
\begin{equation}
T_{\mu}^{\rho}=\frac{i}{4} g^{\rho\nu} \biggl(\bp \gamma_\mu
\nabla_\nu \psi + \bp \gamma_\nu \nabla_\mu \psi - \nabla_\mu \bar
\psi \gamma_\nu \psi - \nabla_\nu \bp \gamma_\mu \psi \biggr) \,-
\delta_{\mu}^{\rho} L_{\rm sp} \label{temsp}
\end{equation}
where $L_{\rm sp}$ in view of \eqref{speq} can be rewritten as
\begin{eqnarray}
L_{\rm sp} & = & \frac{\imath}{2} \bigl[\bp \gamma^{\mu}
\nabla_{\mu} \psi- \nabla_{\mu} \bar \psi \gamma^{\mu} \psi \bigr] -
m_{\rm sp} \bp \psi - F(K)
\nonumber \\
& = & \frac{\imath}{2} \bp [\gamma^{\mu} \nabla_{\mu} \psi - m_{\rm
sp} \psi] - \frac{\imath}{2}[\nabla_{\mu} \bar \psi \gamma^{\mu} +
m_{\rm sp} \bp] \psi
- F(K),\nonumber \\
& = & 2 F_K (I K_I + J K_J) - F = 2 K F_K - F(K). \label{lspin01}
\end{eqnarray}

It can be shown that the spinor field in this case possesses
nontrivial non-diagonal components of the energy-momentum tensor. On
account of that the system of Einstein equations can be written as
\cite{sahalrsbi}

\begin{subequations}
\label{lrsBIEn}
\begin{eqnarray}
\frac{\ddot a_3}{a_3} +\frac{\ddot a_1}{a_1} + \frac{\dot
a_3}{a_3}\frac{\dot a_1}{a_1}&=& \kappa (F(K) - 2 K F_K),\label{11lrsbin}\\
2\frac{\ddot a_1}{a_1}  +  \frac{\dot
a_1^2}{a_1^2}&=&  \kappa (F(K) - 2 K F_K),\label{33lrsbin}\\
\frac{\dot a_1^2}{a_1^2} + 2 \frac{\dot a_3}{a_3}\frac{\dot
a_1}{a_1}&=& \kappa (m_{\rm sp} S + F(K)), \label{00lrsbin}\\
0 &=& \left(\frac{\dot a_3}{a_3} - \frac{\dot
a_1}{a_1}\right)\,A^2,,
\label{13lrsbin}\\
0 &=& \left(\frac{\dot a_1}{a_1} - \frac{\dot a_3}{a_3}\right)\,A^1.
\label{23lrsbin}
\end{eqnarray}
\end{subequations}
where $A^\mu = \bar \psi \gamma^5  \gamma^\mu \psi$ are the
components of the pseudovector.

\section{Solution to the field equations}

In this section we solve the equations obtained in the previos
section. The off-diagonal components of the Einstein equations
\eqref{13lrsbin} and \eqref{23lrsbin} impose the following
restrictions either on the components of the spinor field or on the
metric functions:
\begin{subequations}
\label{reslrsbi}
\begin{eqnarray}
A^2  = 0, \quad A^1 &=& 0,\label{resspinlrsbi} \\
\frac{\dot a_3}{a_3} - \frac{\dot a_1}{a_1} &=& 0.
\label{resmetlrsbi}
\end{eqnarray}
\end{subequations}

From \eqref{resmetlrsbi} we dully find $a_3 = q_0 a_1$ with $q_0$
being some constant. In this case the system can be described by a
FRW model from the very beginning. Hence we don't consider this case
here. We will do it in some of our forthcoming papers on FRW model.

So we consider \eqref{resspinlrsbi} when the restriction is imposed
on the components of the spinor field. Subtraction of
\eqref{33lrsbin} from \eqref{11lrsbin} gives

\begin{eqnarray}
\frac{\ddot a_3}{a_3} - \frac{\ddot a_1}{a_1} +  \frac{\dot
a_1}{a_1} \left(\frac{\dot a_3}{a_3} - \frac{\dot a_1}{a_1}\right) =
0, \label{sub31}
\end{eqnarray}
that leads to \cite{saha2001a}
\begin{eqnarray}
a_1 = D_1 V^{1/3} \exp{\left(X_1 \int \frac{dt}{V}\right)}, \quad
a_3 = (1/D_1^2) V^{1/3} \exp{\left(-2X_1 \int \frac{dt}{V}\right)}.
\label{metricflrsbi}
\end{eqnarray}
with $D_i$ and $X_i$ being the integration constants. Thus we see
that the metric functions can be expressed in terms of $V$.

Our next step will be to define $V$. Combining the diagonal Eistein
equations \eqref{11lrsbin}, \eqref{33lrsbin} and \eqref{00lrsbin} in
a certain way for $V$ we find  \cite{saha2001a}
\begin{eqnarray}
\ddot V = \frac{3 \kappa}{2} \left(m_{\rm sp}\,S +  2 ( F(K) -  K
F_K)\right) V. \label{detvlrsbi}
\end{eqnarray}

Now order to solve \eqref{detvlrsbi} we have to know the relation
between the spinor and the gravitational fields. Using the equations
\begin{subequations}
\label{invlrsbi}
\begin{eqnarray}
\dot S_0  +  {\cG} A_{0}^{0} &=& 0, \label{S0lrsbi} \\
\dot P_0  -  \Phi A_{0}^{0} &=& 0, \label{P0lrsbi}
\end{eqnarray}
\end{subequations}
where we denote $S_0 = S V,\, P_0 = P V$ it can be show that

\begin{equation}
K = \frac{V_0^2}{V^2}, \quad  K = \{I,\,J,\,I+J,\,I-J\}. \label{KV}
\end{equation}

The relation \eqref{KV} holds for  $K = \{J,\,I + J,\,I - J\}$ only
for massless spinor field, while for $K = I$  it holds both for
massless and massive spinor field. In case of $K = I + J$ one can
write $S = \sin{(V_0/V)}$ and $P = \cos{(V_0/V)}$, whereas for  $K =
I - J$ one can write $S = \cosh{(V_0/V)}$ and $P = \sinh{(V_0/V)}$.
In what follows, we will consider the case for $K = I$, setting $F =
\sum_{k} \lambda_k I^{n_k} =  \sum_{k} \lambda_k S^{2 n_k}$, as in
this case further setting spinor mass $m_{\rm sp} = 0$ we can revive
the results for other cases.

Then inserting $F =  \sum_{k} \lambda_{k} I^{n_k} = \sum_{k}
\lambda_{k} S^{2 n_k}$ into \eqref{detvlrsbi} and taking into
account that in this case $S = V_0/V$ we find

\begin{eqnarray}
\ddot V = \frac{3 \kappa}{2} \left[m_{\rm sp}\,V_0 + 2 \sum_{k}
\lambda_k( 1 - n_k) V_0^{2n_k} V^{1 - 2n_k}\right],
\label{detvlrsbinew}
\end{eqnarray}
with the solution in quadrature
\begin{eqnarray}
\int \frac{dV}{\sqrt{3 \kappa \left[m_{\rm sp} V_0 V + \sum_{k}
\lambda_k V_0^{2n_k} V^{2(1-n_k)} + {\bar C}\right] }} =  t + t_0,
\label{quadlrsbi}
\end{eqnarray}
with ${\bar C} $ and $t_0$ being some arbitrary constants.

In what follows we solve the equations for $V$ , i.e.,
\eqref{detvlrsbinew} numerically. But before doing that we write the
solution to the spinor field equations explicitly.

The solutions to the spinor field equations \eqref{speq} in this
case can be presented as
\begin{eqnarray}
\psi_{1,2}(t) = \frac{C_{1,2}}{\sqrt{V}} \exp{\left(-i\int {\Phi}
dt\right)}, \quad \psi_{3,4}(t) = \frac{C_{3,4}}{\sqrt{V}}
\exp{\left(i\int  {\Phi} dt\right)}, \label{psinl}
\end{eqnarray}

with $C_1,\,C_2,\,C_3,\,C_4$ being the integration constants and
related to $V_0$ as
$$C_1^* C_1 + C_2^* C_2 - C_3^* C_3 - C_4^* C_4 = V_0.$$

Thus we see that the metric functions, the components of spinor
field as well as the invariants constructed from metric functions
and spinor fields are some inverse functions of $V$ of some degree.
Hence at any spacetime point where $V = 0$ it is a singular point.
So we consider the initial value of $V (0)$ is small but non-zero.
As a result for the nonlinear term to prevail in
\eqref{detvlrsbinew} we should have $1 - 2 n_k < 0$, i.e., $n_k >
1/2$, whereas for an expanding Universe when $V \to \infty$ as $t
\to \infty $ one should have $1 - 2 n_k > 0$, i.e., $n_k < 1/2$. As
is seen from \eqref{detvlrsbinew}, $n_k = 1/2$ leads to a term that
can be added to the mass term. So without losing the generality we
can consider $n_0 = 1/2$, $n_1 = 0$ and $n_2 = 2$.

In this case we obtain

\begin{eqnarray}
\ddot V = \Phi_1(V), \quad \Phi_1(V) = \frac{3 \kappa}{2}
\left[\left(m_{\rm sp} + \lambda_0\right)\,V_0 + 2 \lambda_1 V - 2
\lambda_2 V_0^{4} V^{-3}\right]. \label{detvlrsbinew1}
\end{eqnarray}
Equation \eqref{detvlrsbinew1} allows the first integral
\begin{eqnarray}
\dot V = \Phi_2(V), \quad \Phi_2(V) = \sqrt{3 \kappa
\left[\left(m_{\rm sp} + \lambda_0\right) V_0 V + \lambda_1 V^2 +
\lambda_2 V_0^{4} V^{-2} + {\bar C}\right]}. \label{1stintlrsbi}
\end{eqnarray}
The solution to the equation \eqref{detvlrsbinew1} can be written in
quadrature as follows
\begin{eqnarray}
\int \frac{dV}{\Phi_2(V)} =  = t + t_0. \label{quadlrsbi1}
\end{eqnarray}

To solve the \eqref{detvlrsbinew1} we should choose the problem
parameters $V_0$, $m_{\rm sp}$, $\kappa$, $\bar C$, $\lambda_k$ as
well as the initial value of $V(0)$ in such as way that does not
leads to
$$ \left(m_{\rm sp} + \lambda_0\right) V_0 V + \lambda_1 V^2 +
\lambda_2 V_0^{4} V^{-2} + {\bar C} < 0.$$

In a recent paper  \cite{sahalrsbi} we considered the case with
$\lambda_0 = \lambda_1 = \lambda_2 = \lambda$. It was shown that in
case of positive $\lambda$ we have an accelerated mode of expansion
of the Universe, while for negative $\lambda$ we have oscillatory
solution.

In this paper we do not perform numerical analysis to obtain
different type of solutions for different values of problem
parameters. In what follows we study the equation
\eqref{detvlrsbinew1} numerically to find the problem parameters
that fit best with the observational data.

\section{Comparison with observations}

In what follows we numerically solve the equation
\eqref{detvlrsbinew}. Let us rewrite this equation as follows:

\begin{equation}
\ddot{V}= A + \sum_{k=1}^3 B_k(1-n_k) V^{1-2n_k}. \label{V-eq}
\end{equation}
Here $A =  \frac{3 \kappa}{2}\left(m_{\rm sp} +
\lambda_0\right)\,V_0$, i.e., the constant term, whereas $B_k =
2\lambda_k V_0^{2n_k}$.

Defining Hubble parameter and red-shift as follows
\begin{subequations}
\begin{eqnarray}
H &=& \frac{1}{3}\frac{\dot{V}}{V} = \frac{\dot a}{a}, \label{Hub}\\
z + 1 &=& \left(\frac{1}{V}\right)^{1/3} = \frac{1}{a}, \label{Zet}
\end{eqnarray}
\end{subequations}
we rewrite \eqref{V-eq} in the form
\begin{eqnarray}\begin{array}{c}
\frac{\partial }{\partial z}\left(\frac{H}{(z+1)^3}\right)^2 =
-\frac{2}{3}\frac{A}{(z+1)^4}-\frac{2}{3}\sum_{k=1}^3
B_k(1-n_k)(z+1)^{6n_k-7}, \label{Zet-Eq} \\ \\ \frac{\partial
}{\partial z}\,t=-\frac{1}{(z+1)^4\sqrt{H}}
\end{array}\end{eqnarray}
with initial values $ t(0)=0$, $H(0)=H_0$.

The foregoing equation allows the following solution
\begin{equation}
H(z) =
\sqrt{\frac{2}{9}A(z+1)^3+\frac{1}{9}\sum_{k=1}^3B_k(z+1)^{6n_k}+C_H(z+1)^6},
\label{HZsol}
\end{equation}
$C_H$ is constant of integration.

The numerical values of the parameters such as spinor mass, power of
nonlinearity etc. are determined by comparing the solutions to
astrophysical observations exploiting the maximum likelihood method
by minimizing the functional

\begin{equation}
\chi^2=\sum\left(\frac{H(z_i)-H_i}{\sigma_{H_i}}\right)^2,
\label{chi}
\end{equation}
where $z_i$, $H_i$ and $\sigma_{H_i}$ were taken from the tables
given in \cite{farook,farook1,zong}.

From \eqref{Zet} it follows that $z\rightarrow\infty$ leads to $a(t)
\rightarrow 0$, i.e. there occurs a space-time singularity. For an
expanding Universe it means $z \rightarrow \infty$ at the time of
{\it Big Bang}. Let us assume that $\dot a |_{z \rightarrow \infty}
= 0$ (soft origin). From  $\dot{a}=\frac{H}{z+1} \rightarrow 0$ we
see that it may happen it $z$ increases faster than $H$ (not
necessarily $H = 0$). Then from \eqref{HZsol} we can conclude that
$A = 0$, $C_H = 0$ and $n_k <1/3 $. The equality $A = 0$ states that
in this case the spinor mass is either zero or compensated by some
constant. Analogical results were found in the papers where dark
energy was simulated by spinor field \cite{sahaAPSS2015}.

The value of $n_k$ is searched using the method of random walk with
the selection of successful steps. At the beginning $n^{(0)}_1=0.1$
and $n^{(0)}_{k+1}=n^{(0)}_k$. At each $m$-th step we  replace
$n^{(m+1)}_k=n^{(m)}_k+0.01\xi$ using standard normal distribution
of random variable $\xi\in N(0,1)$, but under the condition
$\chi^2(n^{(m+1)})<\chi^2(n^{(m)})$.

\begin{figure}[ht]
\centering
\includegraphics[height=60mm]{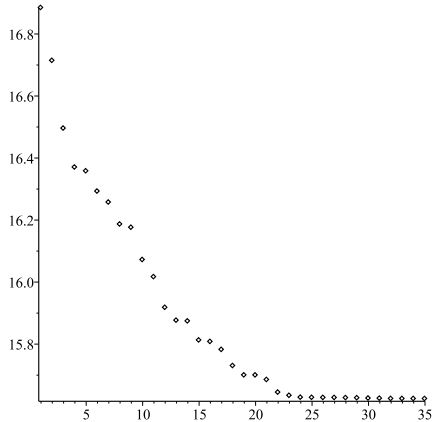} \\
\caption{Dynamics of the process of finding the minimum $\chi^2$}
\label{MMIN}.
\end{figure}

To determine $A$ and $B_k$ at each $m$-th step for fixed $n_k$ the
functional \eqref{chi} is approximately substituted by
\begin{equation}
\tilde{\chi}^2=\sum\left(\frac{H(z_i)^2-H_i^2}{2 H_i
\sigma_{H_i}}\right)^2, \label{chit}
\end{equation}
what is quadratic to coefficients to be determined and the quest of
the minimum is reduced to the solution of linear equations. In Fig.
\ref{MMIN} we have plotted the dynamics of the process of finding
the minimum $\chi$.

The adjusted values of parameters are
\begin{eqnarray}\begin{array}{ll}
n_1=0.1961075852, & B_1=8.033381\cdot 10^6=3\kappa\lambda_1V_0^{2n_1}, \\
n_2=0.1922624717, & B_2=-1.073812\cdot 10^6=3\kappa\lambda_2V_0^{2n_2}, \\
n_3=0.1805648222, & B_3=2.709600\cdot
10^6=3\kappa\lambda_3V_0^{2n_3}.
\end{array}\end{eqnarray}
 At present epoch model time $t=0$, red-shift  $z=0$ and Hubble parameter $H_0=73.5$.

It is known that $z=Ht$ for small $z$, so comparing at red-shift
$z=0.001$ the astronomical time $t^a(0.001)=0.001/H_0$ with model
time $t(0.001)$ we can obtain the age of Universe
$T=[t(\infty)/t(0.001)] t^a(0.001)=15.0\cdot 10^{9}$ years. The
confidence level is found to be $CL=0.92$.

In case of $A\ne 0$ (hard origin), the adjusted values of parameters
are
\begin{eqnarray}\begin{array}{ll}
& A=-39093.17887=\frac{3 \kappa}{2}\left(m_{\rm sp} +
\lambda_0\right)\,V_0, \\
n_1=0.3318171453, & B_1=10.70355153\cdot 10^6=3\kappa\lambda_1V_0^{2n_1}, \\
n_2=0.3122973942, & B_2=-21.81174544\cdot 10^6=3\kappa\lambda_2V_0^{2n_2}, \\
n_3=0.2939425415, & B_3=11.23402881\cdot
10^6=3\kappa\lambda_3V_0^{2n_3}.
\end{array}\end{eqnarray}
 At present epoch model time $t=0$, red-shift  $z=0$ and Hubble parameter $H_0=72.8$,
we can obtain the age of Universe $T=13.7\cdot 10^{9}$ years. The
confidence level is found to be $CL=0.93$.

In Figs. \ref{FHzh_extra} and \ref{FHz_extra}, we have plotted the
$H(z)/(1+z)$ data (32 points) and model prediction (line for
best-fit model) as a function of red-shift (logarithmic scale) for
hard and soft origins, respectively. In Figs. \ref{FHth} and
\ref{FHt} $H(z)/(1+z)$ model prediction as a function of time $t$
(years) has been drawn for hard and soft origins, respectively.
Figs. \ref{FVth} and \ref{FVt} we have demonstrated the evolution of
volume scale $V$ (logarithmic scale) model prediction as a function
of time t (years) for hard and soft origins, respectively.

\myfigures{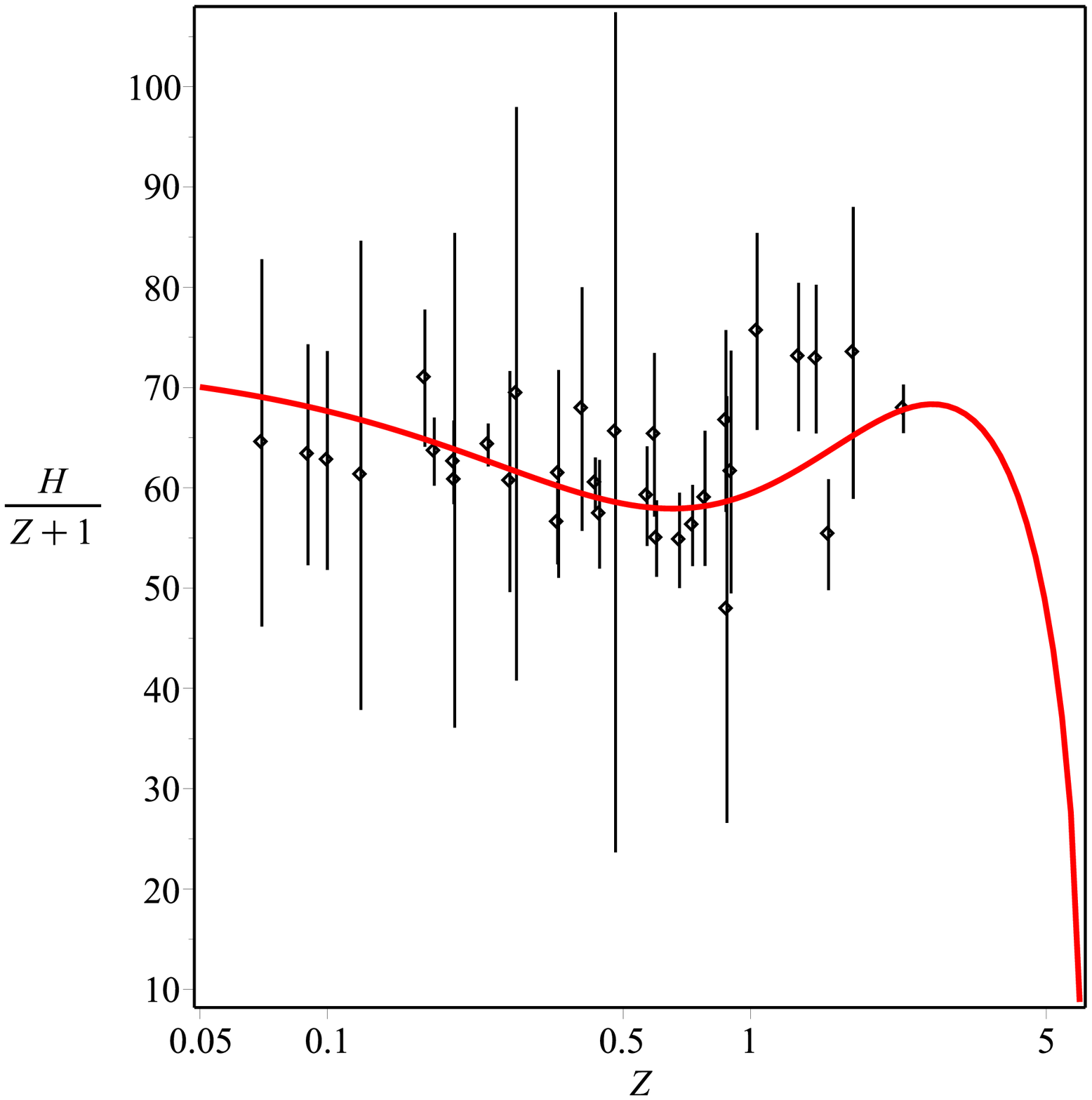}{0.46}{$H(z)/(1+z)$ data (32 points) and model
prediction (line for best-fit model) as a function of red-shift
(logarithmic scale) for hard
origin}{0.45}{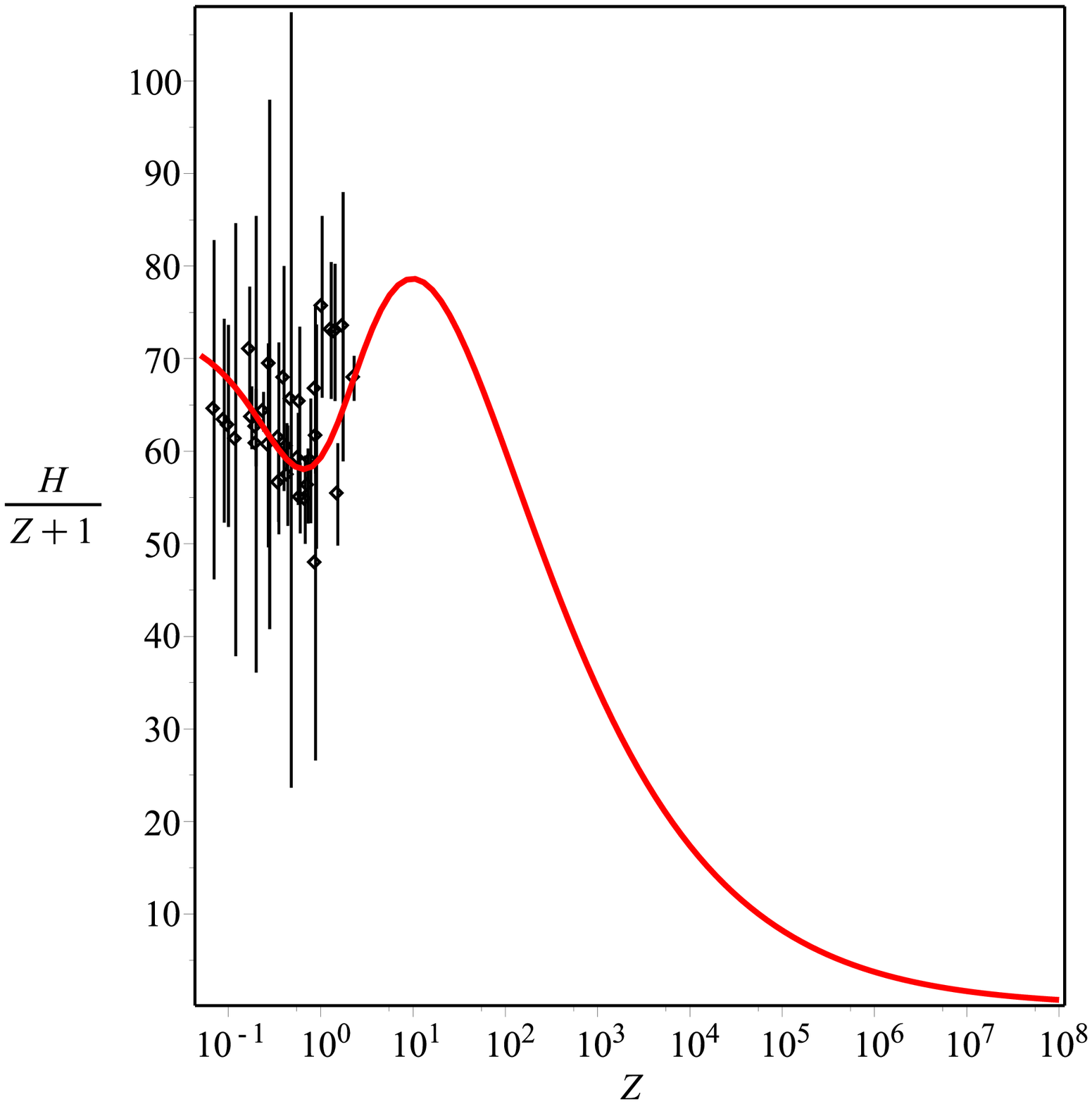}{0.43}{$H(z)/(1+z)$ data (32 points) and
model prediction (line for best-fit model) as a function of
red-shift (logarithmic scale) for soft origin}{0.45}

\myfigures{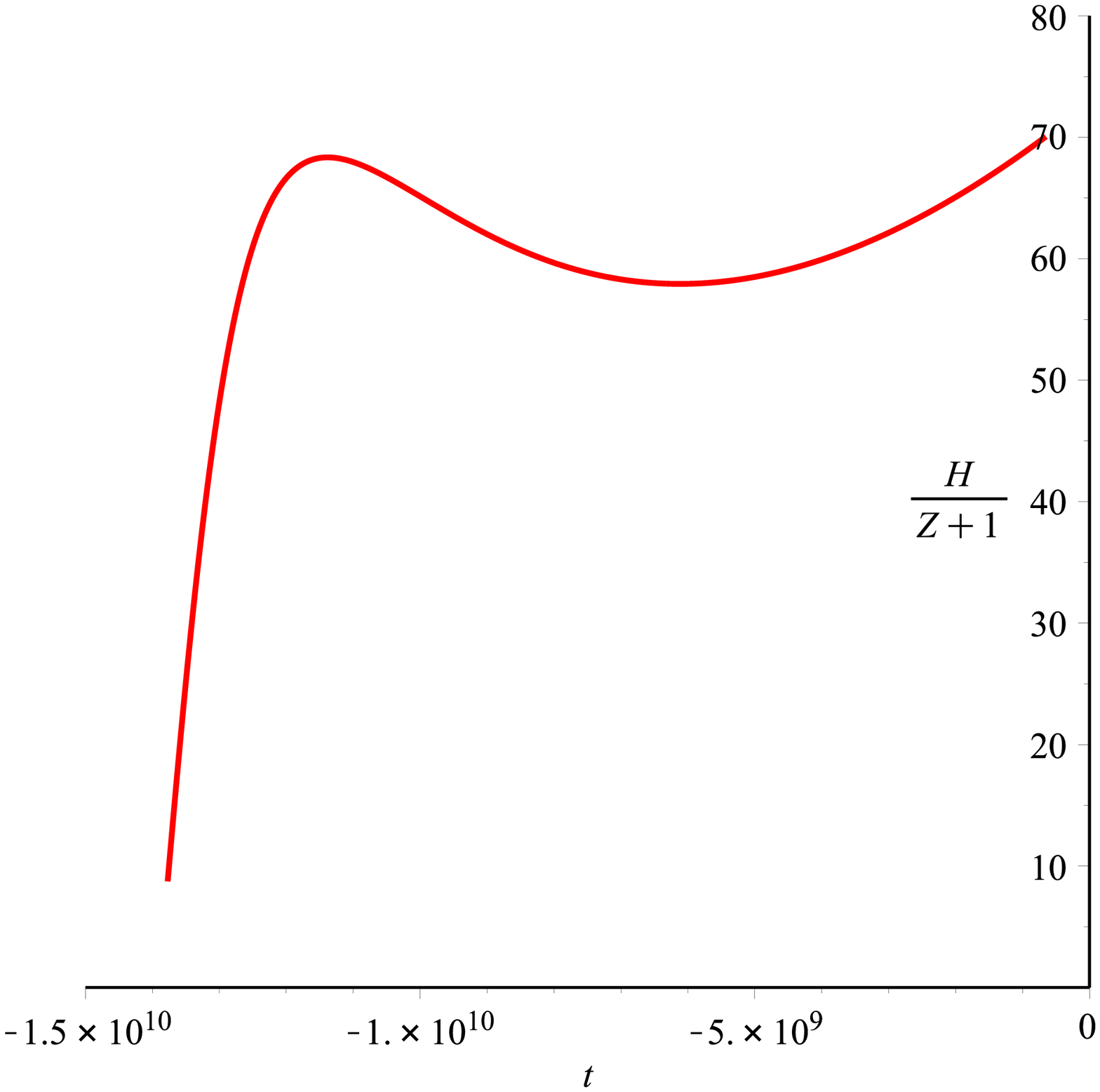}{0.46}{$H(z)/(1+z)$ model prediction as a function
of time $t$ (years) for hard origin}{0.45}{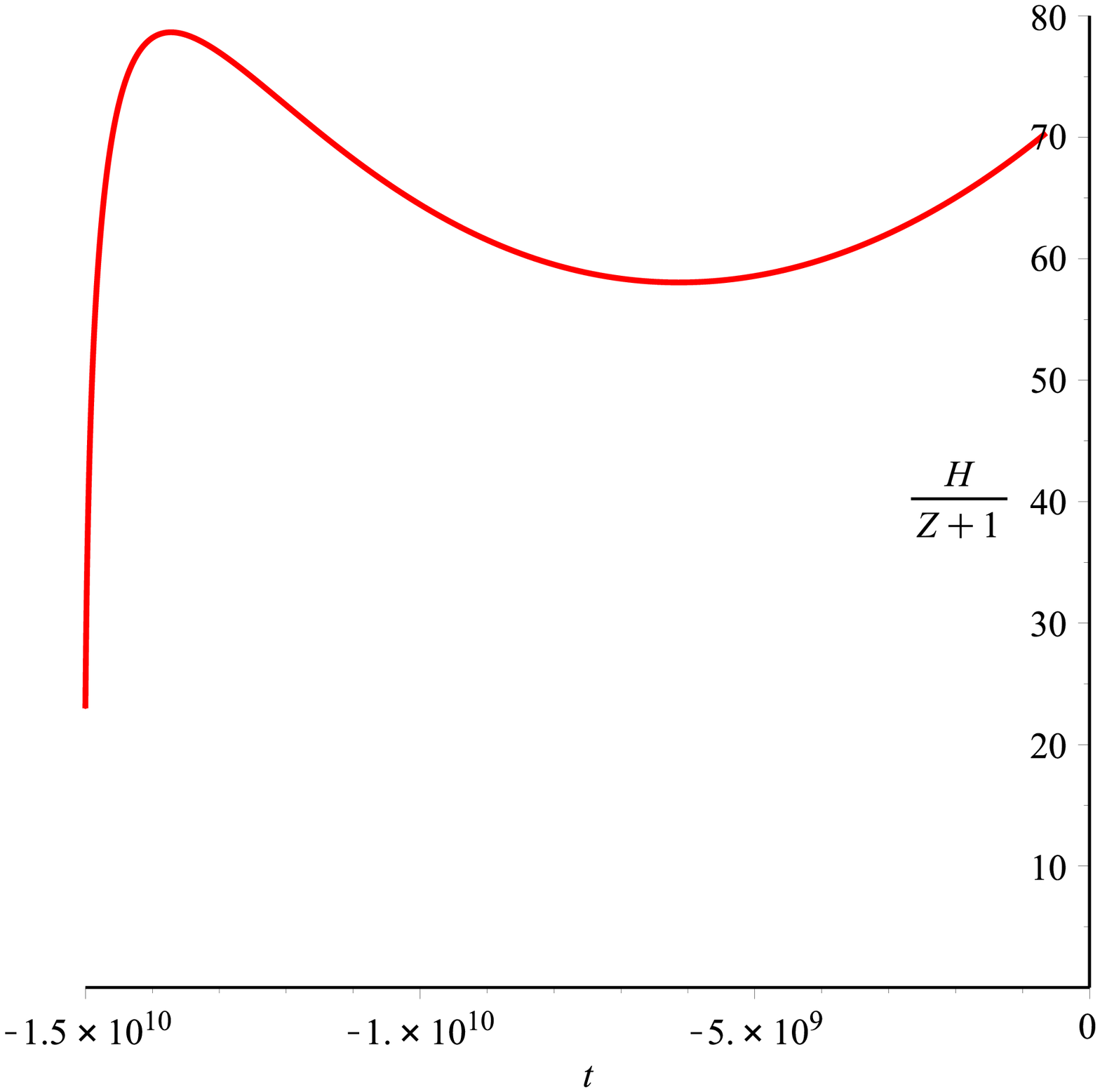}{0.43}{$H(z)/(1+z)$
model prediction as a function of time $t$ (years) for soft
origin}{0.45}

\myfigures{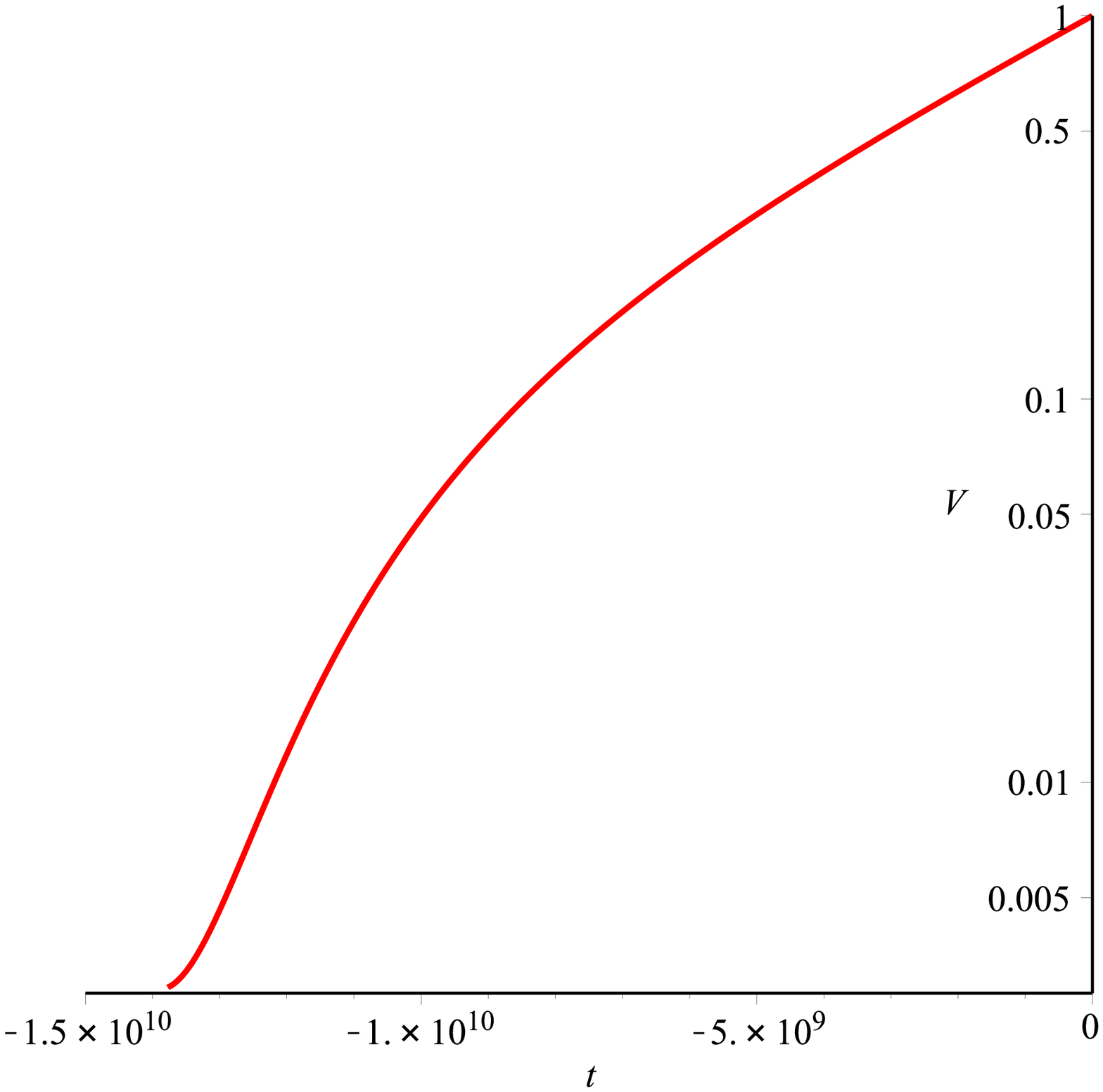}{0.46}{$V$ (logarithmic scale) model prediction as a
function of time $t$ (years) for hard origin}{0.45}{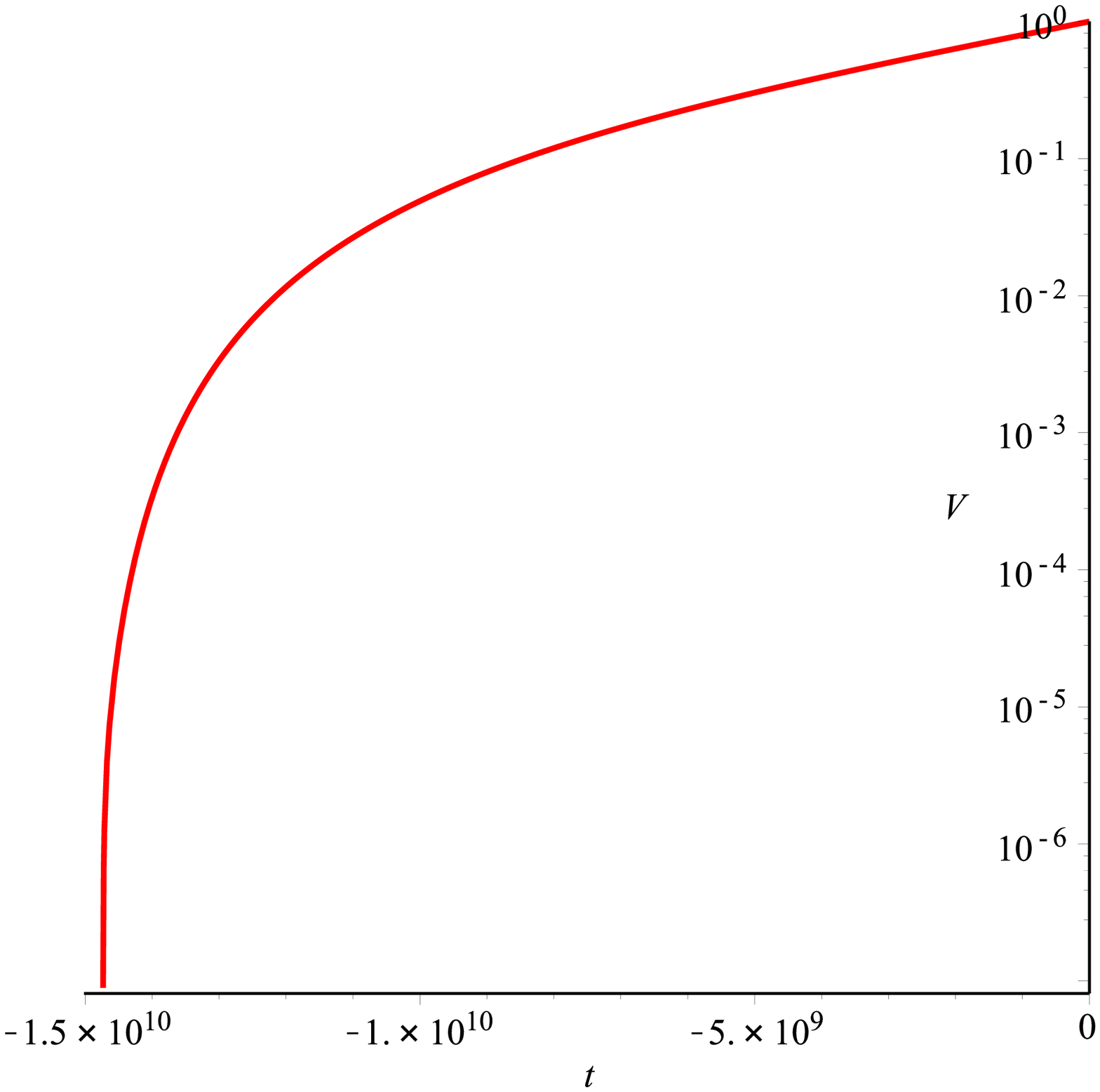}{0.43}{$V$
(logarithmic scale) model prediction as a function of time $t$
(years) for soft origin}{0.45}

\section{Conclusion}

Within the scope of LRS Bianchi type-I model we have studied the
role of spinor field in the evolution of the Universe. Since the
expression for metric functions, components of the spinor field and
invariants constructed from these quantities are in functional
dependence on volume scale $V$ it is important to find and solve the
equation for $V$. In this report we have solved the equation in
question and in doing so we have exploited the astronomical data
available to fix the problem parameters such as coupling constants
etc. that provides the best correspondence with observation.

The assessment of the age of the Universe in case of the soft
beginning of expansion (initial speed of expansion in a point of
singularity is equal to zero) the age was found 15 billion years,
whereas in case of the hard beginning (nontrivial initial speed) it
was found that the Universe  is 13.7 billion years old.

\vskip 5 mm

\noindent {\bf Acknowledgments}\\
This work is supported in part by a joint Romanian-LIT, JINR, Dubna
Research Project, theme no. 05-6-1060-2005/2013.


\end{document}